\documentstyle[aps,prl]{revtex}

\begin{document}
\draft
\title{
Stochastic Resonance in Noisy
Non-Dynamical Systems
}
\author{
J. M. G. Vilar, G. Gomila, and J. M. Rub\'{\i}
}
\address{
Departament de F\'{\i}sica Fonamental, Facultat de
F\'{\i}sica, Universitat de Barcelona, Diagonal 647,
E-08028 Barcelona, Spain 
}
\maketitle
\begin{abstract}
We have analyzed the effects of the addition of external noise to
non-dynamical systems displaying intrinsic noise, and
established general conditions under which
stochastic resonance appears.
The criterion we have found may be applied to a wide
class of non-dynamical systems, covering situations of different nature.
Some particular examples are
discussed in detail.

\end{abstract}

Stochastic resonance (SR)
\cite{Benzi,tri,Ma1,gam,Ma2,neu1,JSP,neu2,%
Moss,Wies,Wiese,array,thre1,grifo,phi4,prl3}
is a phenomenon wherein the addition
of noise to a system enhances its response to a periodic input signal.
This fact is usually characterized by an increase of the output
signal-to-noise ratio (SNR) as the noise level increases.
The main fingerprint of this constructive role played
by noise is the appearance of a maximum in
the SNR at a nonzero noise level,
although recently new manifestations have been found such as
multiple maxima \cite{prl2} or a divergent
SNR with the noise level \cite{prl1}.
In spite of the fact that at the beginning this phenomenon
was  restricted to bistable systems, nowadays
it is known that there are different situations in which SR appears,
as for instance in
monostable, excitable, and non-dynamical systems.

In regards to non-dynamical systems, SR has only been found in
threshold and threshold-like  devices \cite{thre1,thre2} and in the
situation in which
the output of the system consists of a random train
of identical pulses, with the probability of a pulse generation
exponentially depending on the input signal\cite{BV}.
Non-dynamical systems, however, encompass more general situations,
of which many display intrinsic noise whose
effects cannot systematically be neglected.

In this Letter we address precisely the problem of
establishing general conditions under which SR appears
in noisy non-dynamical systems.
We will show that the addition of  external noise to
the periodic input signal may give rise to the enhancement
of the response of the system, therefore implying
the presence of SR.
In this sense, the addition of external noise
helps to overcome the limitations imposed
by the unavoidable intrinsic noise.

A non-dynamical system can be characterized by
its output as a function of given input parameters
which do not depend on the output.
This very general requirement is responsible for
the fact that non-dynamical systems are frequently found in 
many different
scientific areas, encompassing
a wide variety
of situations, including,
to mention just a few: biological membranes,
ionic channels, solid state diodes, quantum dots,
and self-assembled molecular nanostructures.
Usually, the output is a fluctuating quantity
which is characterized by its mean value and variance.
The characteristics of this 
intrinsic noise, to some extent, are determined by
the state of the system,
which in turns depends on the value of the input.
This stochastic behavior is reflected
in the class of systems
described by
\begin{equation}
\label{master}
I(t)=H(V)+\xi(t) \;\; ,
\end{equation}
where $I(t)$ and $V$ are the output and the input, respectively.
The function $H(V)=\left<I(t)\right>$ corresponds to the
deterministic response and $\xi(t)$ represents
the intrinsic noise.
Usually the 
characteristic time scales of $\xi(t)$ are
smaller than any other entering the system. Under this circumstance,
it can be approximated by Gaussian white noise with
zero mean and correlation function
$\left<\xi(t)\xi(t+\tau)\right>=G(V)\delta(\tau)$.
Here, the function
$G(V)=\left<I(t)^2\right>-\left<I(t)\right>^2$
shows the dependence of the
noise on the state of the system.
Eq. (\ref{master}) then describes, in general, the behavior of most
noisy non-dynamical systems. For instance,
in the case of systems consisting of
spike trains with
Poisson statistics, the variance is equal
to the mean, then $G(V)=H(V)$, and $H(V)$ gives the
pulse rate generation.

In order to analyze
how the addition of external noise affects the response
of the system to a slow periodic signal,
we consider the input 
consisting of sinusoidal and random contributions, namely
$V \equiv V(t)=V_s(t)+V_r(t)$.
Here $V_s(t)=\alpha\sin(\omega_0 t)$, with $\alpha$ being
the amplitude and $\omega_0 / 2\pi$ the frequency.
The random term $V_r(t)$ is assumed to be Gaussian noise with
zero mean and correlation function
$\left<V_r(t)V_r(t+\tau)\right>=\sigma^2\exp(-\tau/\tau_F)$,
where $\sigma^2$ defines the noise level and
$\tau_F$  is
the correlation time, which is assumed $\tau_F \ll 2\pi / \omega_0$.
The SNR can be computed from the averaged power spectrum
\begin{equation}
\label{ps}
P(\omega) = 4\int^{2\pi / \omega_0}_0{\omega_0 \over 2\pi} dt
\int^\infty_{0}\left<I(t)I(t+\tau)\right>_{\xi,r} \cos(\omega\tau) d\tau \;,
\end{equation}
which follows from
the correlation function
\begin{eqnarray}
\label{cf}
\nonumber
\left<I(t)I(t+\tau)\right>_{\xi,r} &=& \left<H(V(t))H(V(t+\tau))\right>_r\\
\label{corre} & &+\left<G(V(t))\right>_r\delta(\tau) \;\;.
\end{eqnarray}
Here $\left<.\right>_{\xi,r}$ and $\left<.\right>_r $ indicate average
over both noises and only over $V_r(t)$, respectively.
By considering sufficiently small amplitudes of the input signal
and low noise level,
$H(V)$ and $G(V)$ can be expanded in power series of
$V$, therefore
\begin{equation}
\label{snr1}
\mbox{SNR}={2\pi\alpha^2H^\prime(H^\prime+H^{\prime\prime\prime}\sigma^2)
\over 2G+(4\tau_F {H^{\prime}}^2+{G^{\prime\prime}})\sigma^2} \;\; ,
\end{equation}
where $^{\prime}$ indicates the derivative of the function with respect to
its argument \cite{gentau}.
All the functions have been evaluated at $V=0$.

Concerning the value of $H^{\prime}$, two different situations
can be analyzed.
If $H^\prime\neq0$ and the inequality
\begin{equation}
\label{maxeq}
2H^{\prime\prime\prime}G-4\tau_F{H^{\prime}}^3-H^{\prime}G^{\prime\prime}>0
\end{equation}
holds, the SNR is an increasing function of the noise level,
which reveals that the addition of noise enhances the
response of the system to a weak periodic signal.
As usually occurs, if the SNR decreases for high
noise level, then the SNR presents at least a maximum,
thus indicating the appearance of SR.

As a first example illustrating the applicability of our results,
we will analyze the
particular case discussed in  Ref.  \cite{BV} in our context.
In these circumstances
$G(V) = H(V)=r(0)\exp(V)$,
where $r(0)$ is the pulse rate generation at $V=0$.
Then, Eq. (\ref{maxeq}) indicates that SR appears
for $r(0)\tau_F<1/4$, in agreement with Ref.  \cite{BV,mal}.
Note that  for small amplitudes
of the input signal, following our approach
[Eqs. (\ref{master}), (\ref{ps}) and (\ref{cf})],
the SNR can analytically be computed
giving the same result as the one of Ref. \cite{BV,mal}.

In order to show the  great generality of Eq. (\ref{maxeq}) we will analyze a 
system in which the internal noise does not follow
Poisson statistics, as is the case, for instance, of a model
for electrical conduction which displays saturation.
In this model, $I(t)$ corresponds to the
current intensity and $V$ to an input voltage.
To be explicit, we will consider
\begin{eqnarray}
\label{modelo11} 
H(V)&=& {V+V_0 \over R[1+(V+V_0)^2]^{1/2}}\;\;,\\
\label{modelo12}
G(V)&=& {Q \over [1+(V+V_0)^2]^{1/2}} \;\; ,
\end{eqnarray}
where $R$ and $Q$ are constants.
Here $V_0$ represents a constant bias voltage.
For low values of the total applied voltage ($V+V_0$), the system behaves as 
a linear resistor. For high values, however,
the non-linear terms become important to the extent that the
current intensity saturates.
Semiconductors systems displaying hot electrons effects
constitute a well known example of systems exhibiting this non-linear
behavior \cite{hot2}.
According to our previous analysis [Eq. (\ref{maxeq})], this system
exhibits SR for sufficiently large values of $V_0$.
In Fig. (\ref{fig1}) we have depicted the SNR as a function of the
noise level, for different values of $V_0$.
These results have been obtained by numerically averaging over realizations
of the noise terms.
SR appears for sufficiently large values of $V_0$,
as predicted by the criterion established in  Eq. (\ref{maxeq}).
The occurrence of SR is not exclusive of this particular model.
We have analyzed other models exhibiting saturation \cite{hot1} and
found that SR is a common phenomenon.

Let us now discuss  the case in which $H^{\prime}=0$.
In this situation, the SNR vanishes for $\sigma^2=0$.
Therefore, for low noise level and sufficiently small amplitudes
of the input signal, if $H(V)$ is not symmetric around $V=0$,
the SNR must be an increasing function of $\sigma^2$,
explicitly\cite{snrgen}
\begin{equation}
\label{snr2}
\mbox{SNR}=
{\pi\alpha^2
{H^{\prime\prime\prime}}^2\sigma^4\over 4 G } \;\; .
\end{equation}
One then concludes that, when $H^{\prime}=0$,
SR is always present for sufficiently
small amplitudes of the input signal and its appearance
does not depend on the form of noise $\xi(t)$,  provided that
the noise term is different from zero at $V=0$.
This situation may occur in 
many systems of interest, as for instance
in tunnel diodes \cite{td,hot1},  ionic channels \cite{ic},
and in a general way in any system
exhibiting maxima, minima or inflection points in
the I-V characteristics.

As an example of a system for which $H^{\prime}=0$, that can be solved
analytically, for any value of the amplitude and noise level,
we will consider the case
\begin{eqnarray}
\label{modelo21} 
H(V)&=& V^3 \;\;,\\
\label{modelo22} 
G(V)&=& Q \;\; ,
\end{eqnarray}
where $Q$ is a constant.
The SNR is found to be
\begin{equation}
\label{ana}
\mbox{SNR}=\pi
{
18\sigma^4\alpha^2 + 9\sigma^2\alpha^4 + {9\over8}\alpha^6
\over
2Q+\tau_F( 44\sigma^6 + 54\sigma^4\alpha^2 + {27\over2}\sigma^2\alpha^4 ) }\;\;.
\end{equation}
From the previous result we can elucidate how the SNR behaves in all the
range of values of $\sigma^2$ and $\alpha^2$ .
It is interesting to point out that
for sufficiently small amplitudes of the 
input signal and low noise level ($\alpha \ll \sigma^2 \ll 1$),
Eq. (\ref{ana}) reduces to Eq. (\ref{snr2}),
then the $\mbox{SNR} \sim \sigma^4$ is
an increasing function of $\sigma^2$ and SR appears,
as predicted by our previous general analysis.
For high noise level the behavior is
$\mbox{SNR} \sim \sigma^{-2}$, then it decreases when increasing
the noise level.
Another notable feature is that the SNR is an increasing function
of the noise level when $\tau_F$ is sufficiently small, irrespective
of the value of the amplitude of the input signal.
In Fig. (\ref{fig2}) we have depicted the SNR corresponding to
Eq. (\ref{ana}) as a function of the noise level,
for different values of $\tau_F$.
This figure clearly illustrates the appearance of a maximum in
the SNR at nonzero noise level and the dependence
of the SNR on the correlation time of the input noise,
namely the output of the system is enhanced when decreasing $\tau_F$.

Another interesting situation in which $H^{\prime}=0$ occurs is
when the I-V curve exhibits a plateau, as for instance
in the case of a quantum dot displaying Coulomb gap \cite{qd}.
Under these circumstances, an input of finite amplitude is required before
any effect on the output is produced. Added noise helps
small deterministic signals to reach this finite amplitude.
Therefore, the output may be enhanced giving rise to the appearance of SR.

Finally, in order to elucidate how the addition of noise enhances the
output in the class of systems
we are considering, we have shown in Fig. (\ref{fig3})
two time series
corresponding
to zero and the optimum noise levels for
the explicit case  given through Eqs (\ref{modelo21}) and (\ref{modelo22}).
This figure clearly shows that the addition of an optimal amount of noise
makes the presence of the input signal manifest.

Our analysis has established  general conditions in which
SR emerges in  noisy non-dynamical systems.
In this way, we have shown that  the appearance of SR in non-dynamical
systems is not a particular situation occurring in threshold devices
and spike-train systems with exponential rate generation,
as believed up to now.
On the contrary, SR is a robust phenomenon that may occur
in many different physical, chemical and biological non-dynamical systems.
One aspect we would like to emphasize is  that a simple
expression we have found is sufficient to make
the presence of SR manifest.
This general criterion, which refers to
a wide class of non-dynamical
systems, may directly be applied
to new cases, thereby  extending the
scope and perspectives of SR.

This work was supported by DGICYT of the Spanish Government under
Grant No.  PB95-0881.  J.M.G.V. and G.G. wish to thank Generalitat de
Catalunya for financial support.

\begin{figure}[th]
\caption[f]{\label{fig1}
SNR for the system described through
Eqs. (\ref{modelo11}) and (\ref{modelo12}) as
a function of $\sigma$ for
different voltages: $V_0=0$ (filled circles),
$V_0=0.5$ (empty circles), $V_0=1.0$ (filled squares),
$V_0=1.5$ (empty squares), and $V_0=2$ (filled triangles).
The values of the remaining parameters are $Q=0.1$, $R=1$,
$\tau_F=0.1$, $\alpha=0.32$ and $\omega_0/2\pi=0.1$.
}
\end{figure}

\begin{figure}[th]
\caption[f]{\label{fig2}
SNR [Eq. (\ref{ana})] for the system described through
Eqs. (\ref{modelo21}) and (\ref{modelo22}) as
a function of $\sigma$ for
different correlation times of the input noise.
The values of the parameters are $Q=0.1$, $\alpha=0.3$,
$\tau_F=1$ (continuous line), $\tau_F=0.3$ (dotted line),
$\tau_F=0.1$ (dashed line),
and $\tau_F=0.03$ (dot-dashed line).
}
\end{figure}

\begin{figure}[th]
\caption[f]{\label{fig3}
(a) Input periodic signal
(b) Time evolution of
$\tilde I(t)={1 \over \Delta t}\int^{t+\Delta t}_tI(s)ds$
corresponding to Eqs. (\ref{modelo21}) and (\ref{modelo22}),
with $\alpha=0.4$, $\omega_0/2\pi=1$,
$Q=10^{-5}$, $\Delta t = 2.4\times10^{-4}$, $\tau_F=10^{-4}$
and $\sigma^2=0$.
(c) Same situation as in (b) but $\sigma^2=0.4$.
}
\end{figure}


\begin{references}
\bibitem{gentau}
The previous results can directly be generalized to the case
in which the correlation function of the external noise is not
exponential, i.e.
$\left<V_r(t)V_r(t+\tau)\right>=\sigma^2f(\tau)$, with $f(0)=1$,
by replacing $\tau_F$ by $\int^\infty_0 f(\tau) d\tau$.
\bibitem{mal}
Note that there is a numerical factor missing in Eqs. (7)
and (8) of Ref. \cite{BV}.
If this factor is taken into account, the condition
for the appearance of SR obtained in  Ref. \cite{BV}
coincides with our condition.
\bibitem{snrgen}
For the case in which the $n$ first odd derivatives of $H(V)$
vanish, the corresponding SNR is found to be
$\mbox{SNR}=2\pi{\alpha^2 \over 2G}\left[H^{(2n+1)}
{1 \over n!}\left({1 \over 2}\sigma^2\right)^n\right]^2$, where $H^{(2n+1)}$
is the $(2n+1)-{th}$ derivative.
\end{references}
\end{document}